\documentclass[12pt, aps,prl,reprint]{revtex4-1}

\usepackage{graphicx}
\usepackage{amsmath}
\usepackage{amsthm}
\usepackage{amsfonts}
\usepackage{epstopdf}
\usepackage[usenames,dvipsnames,svgnames,table]{xcolor}

\begin{document}

\author{Antonio Caretta}
\email{antonio.caretta@elettra.eu}
\author{Michiel~C. Donker}
\author{Alexey~O. Polyakov}
\author{Thomas~T.~M. Palstra}
\author{Paul~H.~M. van Loosdrecht}
\email{pvl@ph2.uni-koeln.de}
\affiliation{Zernike Institute for Advanced Materials, University of Groningen, Nijenborgh 4, 9747 AG Groningen, The Netherlands}
\title{Photo-induced magnetization enhancement in two-dimensional\\ weakly anisotropic Heisenberg magnets}

\begin{abstract} 
By comparing the photo-induced magnetization dynamics in simple layered systems we show how light-induced modifications of the magnetic anisotropy directly enhance the magnetization. 
It is observed that the spin precession in ${(CH_3NH_3)_2CuCl_4}$, initiated by a light pulse, increases in amplitude at the critical temperature $T_C$. 
The phenomenon is related to the dependence of the critical temperature on the axial magnetic anisotropy. 
The present results underline the possibility and the importance of the optical  modifications of the anisotropy, opening new paths toward the control of the magnetization state for ultrafast memories.

\end{abstract}

\maketitle

Fast and efficient control of the magnetic state of a material requires the understanding of non-equilibrium processes happening in the picosecond time regime. 
The importance of studying the magnetic phenomena with femtosecond time-resolution was first demonstrated by Beaurepaire \emph{et al} \cite{beaurepaire96}. 
In this work it was found that ferromagnetic Ni could be demagnetized with ultrashort optical pulses within one picosecond. 
Shortly after this observation, a number of experiments showing optical manipulation of the magnetization were published \cite{nurmikko99,koopmans02,kimel04}. 
The photo-generation of coherent spin precession as well as the possibility of controlling the spin precession direction via the inverse Faraday effect \cite{kimel05} were demonstrated. 
These investigations lead finally to the spectacular observation of all-optical magnetization switching in GdFeCo \cite{stanciu07}. 
Recently the field of ultrafast magnetization, ``femtomagnetism'' \cite{kimel10}, started to address increasing attention towards other possible ultrafast switches: 
propagating acoustic phonons \cite{bigot12,temnov13} and spin waves \cite{satoh12}, or spin polarized electron transport \cite{eschenlohr13} and spin transfer torque \cite{nemec12}. 
All these phenomena are essentially non-equilibrium processes, and the ability to control the magnetic state of a material by optical pulses has obvious potential in applications. 
The generation of coherent spin precession is a possible method to control the magnetic state of a system. \\
A spin precession can be initiated by an optical pulse via different mechanisms. 
Ultrafast demagnetization in metallic thin films can for instance induce a fast ($<10$~ps) spin reorientation \cite{koopmans02}\footnote{The change of the shape anisotropy is the non-equilibrium condition for precession: the reorientation of the equilibrium axis is as fast as the demagnetization in the material.}.  
Also the use of circularly polarized optical pulses can efficiently initiate a precession \cite{kimel05,bossini14}. 
The first method relies on fast demagnetization rates, and the second on the transparency of the material and strong spin-orbit coupling. 
In order to investigate the other mechanisms of spin reorientation, we measured two insulating magnetic materials, characterized by a unique range of parameters, low demagnetization rate and nearly quenched spin-orbit coupling. 
The organic-inorganic hybrid materials realize these requirements, being additionally versatile, tunable and multifunctional \cite{mitziprogr,hmtopic,caretta14}.
In particular it would be interesting to investigate the effect of non-equilibrium modifications of the magnetic parameters like the single-ion magnetic anisotropy. 
The modifications of the magnetic anisotropy show up, phenomenologically, at time scales only limited by the magnetic response of the system, the spin precession period. 
These processes might be extremely relevant for the understanding all-optical magnetization control, but few experiments address this issue. \\

In this Letter we report on optical time-resolved Faraday rotation (FR) experiments on simple magnetic systems, the layered organic-inorganic hybrids (OIHs). 
The magnetic field and temperature dependence of the light-induced precession
is studied in two model systems: ferromagnetic ${(CH_3NH_3)_2CuCl_4}$ (MACuCl), and antiferromagnetic ${(CH_3CH_2NH_3)_2CuCl_4}$ (EACuCl). 
The observed difference in response of the two materials strongly suggests that, even in the low-fluence regime, the magnetization enhancement of the spin precession observed in MACuCl is due to a non-equilibrium increase of the axial magnetic anisotropy. 
Before presenting the main results we will first review some basic magnetic properties of the organic-inorganic hybrids. 

\begin{figure}
	\centering
		\includegraphics[width=\columnwidth]{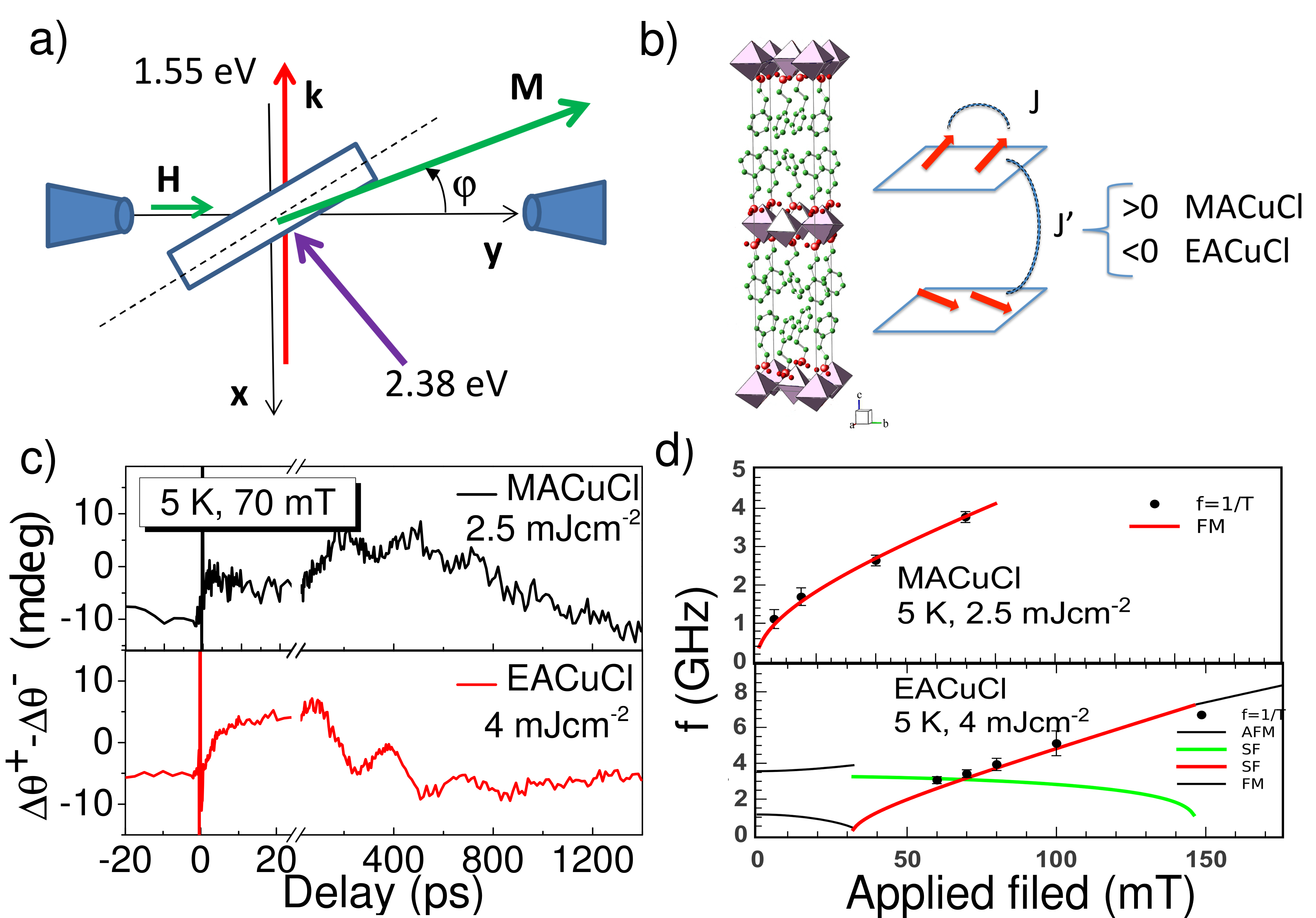}
\caption{(Color online) a) Experiment scheme. 
b) Crystal structure of EACuCl and exchange constants summary.
c) Magnetization dynamics in MACuCl and EACuCl at 70~mT and 5~K. 
d) Precession frequency measured from the Faraday rotation in MACuCl (up) and EACuCl (down). 
The $f(H)$ curves represent the resonance frequency in the antiferromagnetic (AFM), the spin-flip (SF) and the ferromagnetic (FM) phases (EACuCl: more details can be found in Ref.~\cite{tanimoto76}).}
	\label{fig:dqH}
\end{figure}

The layered Cu-based organic-inorganic hybrids here studied are nearly ideal 2-dimensional Heisenberg ferromagnets and have been widely studied in the 70's \cite{dejongh74}. 
The structure is constituted by corner sharing CuCl$_6$ octahedra, forming the inorganic layers, and bilayers of organic cations, attached to the octahedra by the NH$_3$ heads \cite{alex2011,caretta14} (fig. \ref{fig:dqH}b). 
The proper Hamiltonian to describe the single layer magnetism is \cite{dejongh74}:
\begin{equation}\label{eq:ham}
{\cal H} = -2 J \sum_{i>j} \vec{S}_i \cdot \vec{S}_j   + \sum_i \left( K S_{yi}^2 +D S_{zi}^2 \right),
\end{equation} 
where $\vec{S}_i$ is the spin of the $i$-th site, $S_{yi}$ and $S_{zi}$ respectively the $y$ and $z$ components of the spin, and $K>0$ and $D>0$ express the magnetic anisotropy. 
The intralayer exchange $J$ is ferromagnetic due to cooperative Jahn-Teller ordering \cite{khomskii_kcuf}, with $J/k\sim$20~K. 
The exchange interaction between the layers is expressed with $J'$ (fig. \ref{fig:dqH}b), and the Hamiltonian is modified by an additional term accounting for the other layers (see ref.~\cite{jonghstanley76}). 
The interlayer exchange $J'$ is up to five orders of magnitude smaller than $J$, and determines the overall three dimensional magnetic character of the compound: while MACuCl orders ferromagnetically at 8.9~K, EACuCl orders antiferromagnetically at 10.2~K \cite{dejongh74}. 
It is well known that 2-dimensional Heisenberg magnets do not order at any finite temperature \cite{mermin66}. 
The only possible cause of ordering in Cu-based OIHs is the presence of finite interlayer exchange $J'$, breaking the 2-d character, or axial magnetic anisotropy $H_A^{Ising}\equiv 2KS/g\mu_B$ \footnote{For $D\gg K$ the layer has planar anisotropy, and for $D \simeq K$ axial anisotropy. 
The magnetic system can be described respectively with a XY model or a Ising model, thus the names $H_A^{XY}$ and $H_A^{Ising}$.}. 
The preferential axis breaks the continuous symmetry and the system behaves similarly to an Ising 2-d magnet, thus orders at a finite temperature \cite{mills88,mills91,petrakovskii88}. 
The largest of the deviations from the ideal situation causes magnetic ordering. 
In table \ref{tabmat} the magnetic parameters describing  MACuCl and EACuCl are shown, with the use of the intralayer effective field $H_E=2zJS/g\mu_B$ ($z$ is the number of neighbors in the layer). 
\begin{table}
\begin{tabular}{lccccc}	
\hline\hline
	&	$T_C$(K) 	& $J/k$(K)		& 	$J'/J$	& $H_A^{Ising}/H_E$	& 	$H_A^{XY}/H_E$	\\
\hline
MACuCl	&8.9&19.2&	$+5.5\times10^{-5}$ &1.6$\times10^{-4}$&3$ \times 10^{-3} $\\
EACuCl &10.2&18.6&$-8 \times 10^{-4}$&1.6$ \times 10^{-4}$&3$ \times 10^{-3} $\\
\hline\hline
\end{tabular}
\caption{Parameters characterizing the magnetic phase of MACuCl and EACuCl\cite{dejongh74,annethesis}: 
the critical temperature $T_C$, the intralayer ferromagnetic exchange $J/k\sim20$~K, the interlayer exchange $J'$. 
The planar and the axial anisotropy effective fields, respectively $H_A^{XY}\equiv 2DS/g\mu_B$ and $H_A^{Ising}\equiv 2KS/g\mu_B$, are normalized by the intralayer effective field $H_E=2zJS/g\mu_B$, where $z=4$ is the number of neighbors in the layer.} \label{tabmat}
\end{table}
Specifically for MACuCl: $H_A^{Ising}/H_E > |J'/J|$. 
Vice versa for EACuCl: $H_A^{Ising}/H_E< |J'/J|$ (see table \ref{tabmat}). 
It is very important to recall the dependence of the critical temperature $T_C$ on the magnetic anisotropy, when $H_A^{Ising}/H_E> |J'/J|$ \cite{mills88,mills91,petrakovskii88}:
\begin{equation} \label{eq:tc}
T_C=\frac{4\pi J}{ln(J/K)}.
\end{equation}
It is clear that modifications of $K$, thus $H_A^{Ising}$, will only affect MACuCl, as instead in EACuCl the critical temperature is, in first approximation, only a function of $J'$ \cite{jonghstanley76}. 

The samples are grown from solution in an analogy to PEACuCl (see ref. \cite{alex2011, arend}). 
Sample thickness is approximately 150~$\mu$m. 
Magnetic properties are probed by Faraday rotation, at a fixed frequency of 1.55~eV (800~nm). 
Time-resolved experiments are performed using a 1 kHz Ti:Sapphire based amplified laser system (Hurricane, Spectra-Physics). 
The 2.38~eV (520~nm) pump is generated using a non-linear optical parametric amplifier (Light Conversion). 
The 520 nm pulses are optimally tuned to the Cu-Cl charge transfer edge \cite{heygster1977,caretta13}. 
The pump fluence used in the experiments ranges from $2.5$~mJcm$^{-2}$ to $4$~mJcm$^{-2}$, corresponding to a temperature increase of approximately 1~K (Supplementary data). 
The time resolution of the experiment is approximately 150~fs \footnote{The cross-correlation had been measured on a ZnSe non-linear crystal and on the sample itself.}. 
Magnetic measurements are performed in tilted-Voigt geometry, with applied magnetic field along $\hat{y}$ perpendicular to the light propagation along $\hat{x}$ (fig. \ref{fig:dqH}a). 
The linearly polarized pump electric field is oriented along $\hat{z}$, parallel to the crystal optical axis. 
The samples are rotated of $\varphi_0 \sim45^\circ$ around the $\hat{z}$ axis. 
In this configuration the  FR is expressed by
\begin{equation}
\theta \propto \vec{k}\cdot \vec{M} \propto |M| sin(\varphi),
\end{equation}
where $\varphi$ is the angle between the magnetization $\vec{M}$ and $\hat{y}$. 
The experimental data are represented as $(\Delta \theta^+-\Delta \theta^-)$, where $\Delta \theta^+$ is the FR signal measured at positive applied field $+H$ and $\Delta \theta^-$ at $-H$. 
It is easy to note that (Supplementary data)
\begin{equation} \label{eq:dq}
\Delta \theta^+-\Delta \theta^- \propto  \left( \frac{\Delta M}{M_0} +m(T)\Delta\varphi \right).
\end{equation}
where $M_0$ is the saturated magnetization measured at 0~K, $\Delta M$ and $\Delta \varphi$ are the pump induced changes of magnetization and orientation, $m(T)\equiv M(T)/M_0$ is the normalized magnetization. 
It is important to note that at low temperatures, where $m(T)\sim 1$, we are sensitive to changes of $\varphi$ while close to $T_C$, where $m(T) \sim 0$, modification of the magnetization absolute value dominates. 
Because of this at high temperatures variations of magnetization orientation should be enormous to compensate for the low value of $m(T)$.

The magnetization dynamics, in both MACuCl and EACuCl, is characterized by two time regimes. 
The time dependence of the Faraday rotation at 70~mT and 5~K, in respectively MACuCl and EACuCl, is shown in figure \ref{fig:dqH}c. 
At early time-scales, up to 20~ps after the pump arrival, the magnetization increases by approximately $0.5-1$\% with a rise time $\tau<10$~ps (Supplementary data). 
This increase does not depend on the applied field or on the pump polarization. 
Later times are instead characterized by oscillations, having an amplitude of the same order as the initial increase and a period strongly dependent on the applied magnetic field. 
The oscillation frequency $f=1/T$ of MACuCl and EACuCl (top and bottom of fig.~\ref{fig:dqH}d) at 5~K is shown as a function of the applied field. 
The frequencies are obtained from the maxima of the oscillations, by averaging the periods. 

The magnetic field dependence of the oscillations matches the precession frequency measured in equilibrium conditions. 
The increase of the precession frequency $f$ as a function of the applied field $H$ in MACuCl can be described by 
$f= \gamma \frac{g}{4\pi}\sqrt{H(H+H_A)}$\cite{koopmans02},
where $\gamma$ is the gyromagnetic factor, $g$ the Landau splitting factor and $H_A$ the equilibrium magnetic anisotropy. 
For MACuCl an estimate for $H_A\sim 200$~mT is found, consistent with the static anisotropy field $H_A^{XY}$ reported in literature (see also Table \ref{tabmat}). 
For EACuCl the precession frequency matches the antiferromagnetic resonance frequency reported in literature \cite{tanimoto76}, and shown in figure \ref{fig:dqH}d. 
It should be noted that, in both MACuCl and EACuCl, the amplitude of the precession is comparable to the initial increase of the magnetization and weakly depends on the applied magnetic field. \\

\begin{figure}
	\centering
		\includegraphics[width=\columnwidth]{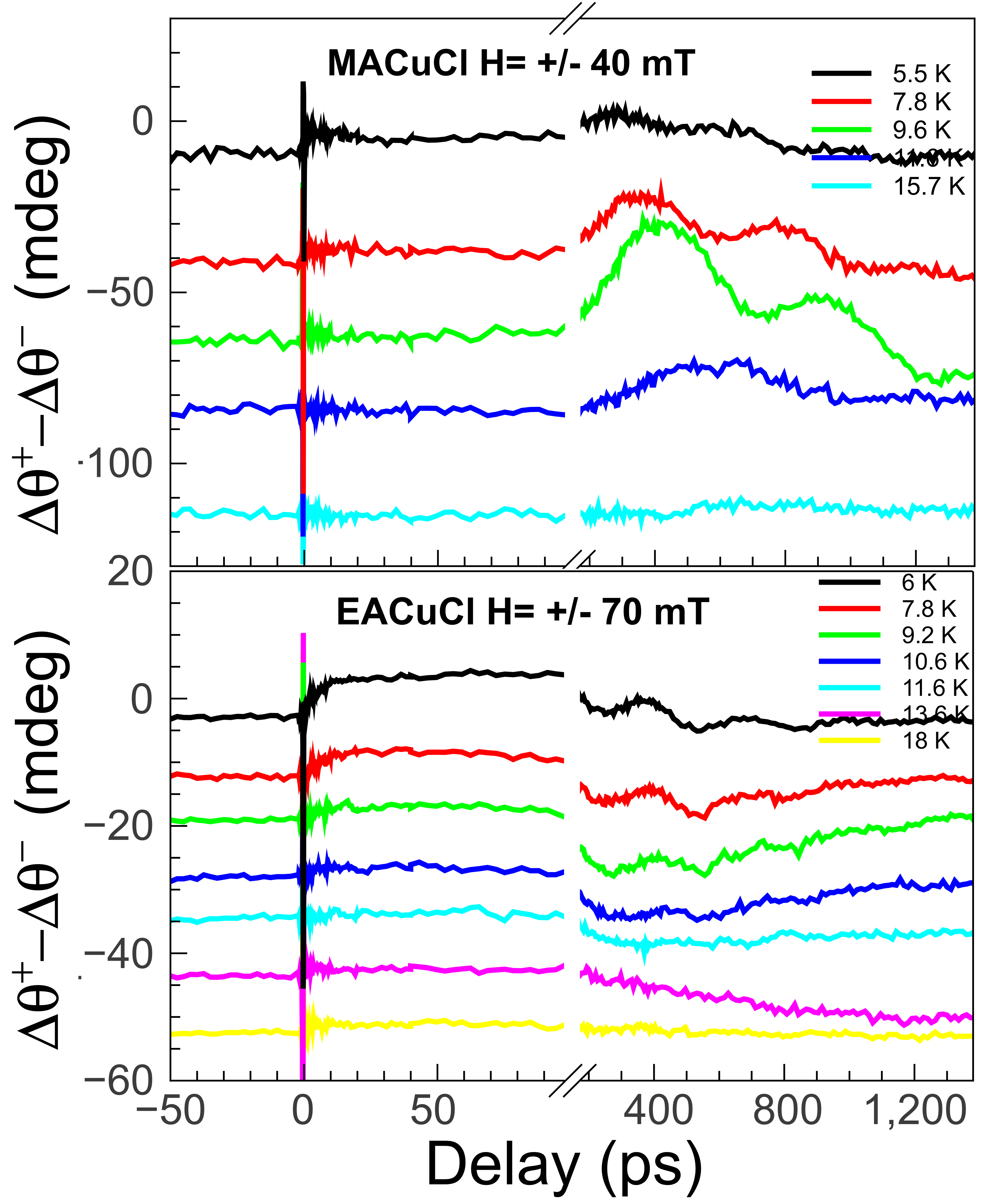}
\caption{(Color online) Temperature dependence of the magnetization dynamics in MACuCl (top, $H=\pm40$~mT) and EACuCl (bottom, $H=\pm70$~mT).
}
	\label{fig:dqT}
\end{figure}

\begin{figure}
	\centering
		\includegraphics[width=.95\columnwidth]{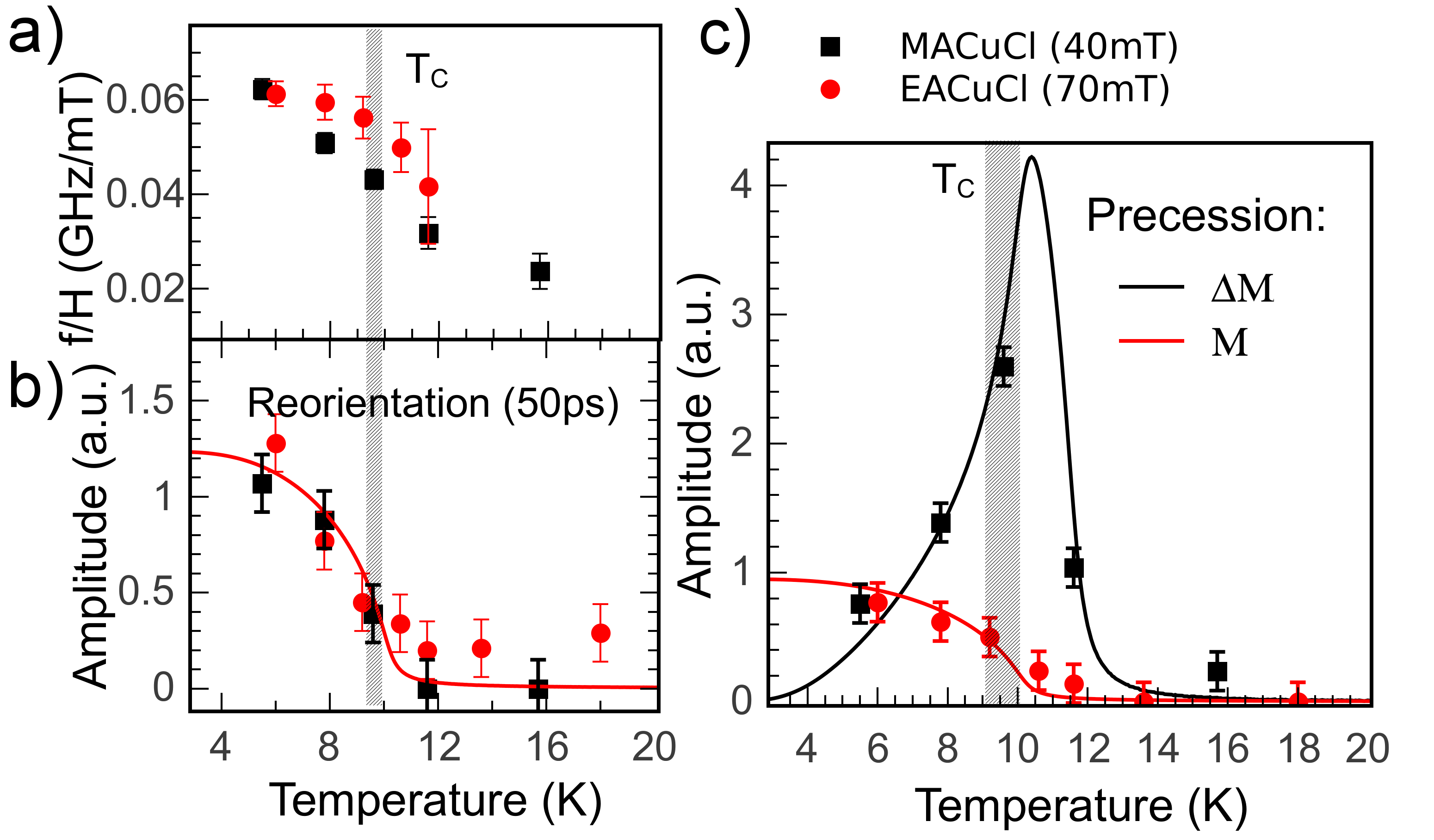}
\caption{(Color online) a) Precession frequency (normalized by the applied field); b) reorientation amplitude (at 50~ps delay) and c) precession amplitude in MACuCl and EACuCl as a function of temperature. 
Note that the amplitudes are expressed as $(\Delta \theta^+-\Delta \theta^-)/(\theta_0^+-\theta_0^-)$. 
$M$ and $\Delta M$ are respectively the magnetization and its derivative calculated at an applied field comparable to the experiment.
}
	\label{fig:afa}
\end{figure}

While the low temperature magnetic field behavior of MACuCl and EACuCl is analogous, the response of the two systems is radically different when crossing the critical temperature. 
The temperature dependence of the magnetization dynamics in MACuCl (top) and EACuCl (bottom) is shown in figure \ref{fig:dqT}. 
The measurements are performed at fixed applied field. 
It can be easily observed that, while in EACuCl the magnetic signal simply disappears with increasing temperature, in MACuCl the precession amplitude grows to a maximum at 9.6~K, close to $T_C$. 
In both compounds the initial increase of the magnetization, between 10~ps and 100~ps, smoothly decreases with increasing temperature. 
Also the precession frequency, weakly decreasing at higher temperatures, seem to be comparable in the two materials. 
In order to quantify these observations, we estimate the initial reorientation amplitude by plotting the increase of $\Delta \theta^+-\Delta \theta^-$ at 50~ps; 
the amplitude and the frequency of the precession are estimated from the oscillation maxima. 
Note that the precession frequencies are normalized by the external magnetic field $H$, as $f/H$. 
The resulting graphs are shown in figure \ref{fig:afa}, where MACuCl and EACuCl are compared. 
As expected the behavior of the two compounds is analogous when looking at the normalized precession frequency $f/H$ and the initial reorientation (fig. \ref{fig:afa}a and \ref{fig:afa}b). 
The initial reorientation scales as the spontaneous magnetization $m(T)$, thus sharply decreases as $T_C$ is approached from below. 
It should be noted that this is an indication that the initial dynamics is dominated by the reorientation of the magnetization, rather than a change of its absolute value. 
In fact, as given in eq. \ref{eq:dq}, the angular term $\Delta \varphi$ is scaled by  $m(T)$ (red curve in fig. \ref{fig:afa}b SI\footnote{$m(T)$ is the magnetization calculated in presence of a small magnetic field, as in the experiments}).
The precession frequency shows a gentle redshift with increasing temperature, in line with temperature dependence of the spin resonance. 
The most striking difference between MACuCl and EACuCl is the amplitude of the oscillation (fig. \ref{fig:afa}c), which decreases to zero at $T_C$ in the latter and grows of a factor of 3, when compared to the value at 4~K, in the former. 

The enhancement of the magnetization absolute value observed in MACuCl, and peaked at $T_C$, can be explained by a non-equilibrium increase of the axial anisotropy. 
We assume that the pump pulses produce the same electronic and structural modifications in both compounds at short (50 ps) timescales. 
This assumption is reasonable since both materials are, in first approximation, a collection of loosely coupled ferromagnetic layers. 
If one additionally notes that the layers are essentially the same (see table \ref{tabmat}), the response to the optical excitation, at least within the layers, must be analogous in the two systems. 
As shown in equation \ref{eq:dq} the FR signal is the sum of an angular term, proportional to the magnetization $m(T)$, and a term proportional the relative changes of the absolute magnetization $M$. 
Close to $T_C$ $m(T)\Delta \varphi\sim 0$, and the observed changes of FR are dominated by changes of the absolute magnetization $M$. 
Clearly, although the two systems initially behave similarly (fig. \ref{fig:afa}b), the absolute magnetization of MACuCl increases in proximity of $T_C$, as shown by the precession amplitude (fig. \ref{fig:afa}c). 
The Ising anisotropy is the only layer parameter which, in first approximation, can specifically affect the magnetization dynamics in MACuCl, leaving EACuCl unperturbed. 
In fact, assuming sizeable changes of $H_A^{Ising}$ (thus changes of $K$ in equation \ref{eq:tc}), only the critical temperature of MACuCl would have notable modifications. 
Changes of the critical temperature value are enhanced at $T_C$, where larger is the slope of the spontaneous magnetization. 
This is indeed observed in MACuCl, and it is easy to quantify the effect.

The increase of magnetization in MACuCl, due to an increase of the anisotropy $H_A^{Ising}$, can be easily estimated from equation \ref{eq:tc}. 
The increase of $T_C$ in MACuCl caused by a increase of the anisotropy $K+\Delta K$ is:
\begin{equation} \label{eq:tc2}
T_C(K+\Delta K)
	 	=T_C\left(1+\frac{1}{ln(J/K)}\frac{\Delta K}{K}\right).
\end{equation}
In the specific case of MACuCl we have:
$$
\Delta T_C \sim \frac{T_C}{10} \frac{\Delta K}{K}\sim \frac{\Delta K}{K}.
$$
Additionally the magnetization, because of an increase of $T_C$, is enhanced and the effect is larger in particular at the critical temperature, since:
$$
\frac{\Delta M}{M_0}=\left | \frac{\partial m}{\partial T}\right | \Delta T_C \sim \left |\frac{\partial m}{\partial T}\right | \frac{\Delta K}{K}.
$$
In fig. \ref{fig:afa}c $\Delta M \propto \frac{\partial m}{\partial T}$ is calculated from the magnetization curve $M$ in applied field (supplementary info) and respectively shown in black and red curves.
Given that, close to $T_C$, $\frac{\Delta M}{M_0} \sim 0.03$ in MACuCl and that the maximum of $\frac{\partial m}{\partial T}$ is approximately 0.4, we obtain an increase of the anisotropy of $\frac{\Delta K}{K} \sim 10$~\%.

Such a small increase of the magnetic anisotropy, which perfectly describes the observations, does not influence the values of the precession frequencies measured at 5~K. 
In particular it has no role for the magnetization dynamics in EACuCl, and the precession amplitude simply scales as the magnetization $m(T)$, reducing to zero at $T_C$ as indeed observed. 

We now briefly discuss the possible microscopic origin of the anisotropy enhancement. 
It should be noted that, because $H_A^{Ising}$ is so small in the hybrids, it is rather difficult to unambiguously determine the origin of the enhancement. 
Given this, in accordance with \cite{temnov13}, the simplest explanation is that the photo generated acoustic phonons distorts the lattice and, due to magneto-elastic coupling, the axial anisotropy is increased. 
Acoustic phonons are nearly instantaneously created by the pump pulses, and generate a long lasting (400 ps) anisotropic distortion, as indeed observed in our FR experiments.  

The effect of anisotropy induced magnetization enhancement can be increased by using multilayer structures based on thin permalloy layers and dielectric spacers. 
Permalloy materials are well known for the high critical temperature and the high magnetic permeability. 
These properties result from respectively a large magnetic exchange $J$ and a low single ion anisotropy $K$. 
As a consequence permalloy materials are good approximations of Heisenberg systems, with critical temperatures up to $T_C=k_BJ\sim$1000~K. 
As was outlined before, when thin films are formed, the critical temperature is expected to decrease, because of the increasing 2-dimensional character of the system \cite{wills94}. 
For a thin enough system the equation \ref{eq:tc} is applicable and the layer ordering temperature is largely reduced. 
By using, as interlayer spacer, a material that can induce stress (e.g. piezoelectric) when an external stimulus is applied, the magnetization can be largely enhanced. 
For instance one monolayer of Ni$_{83}$Fe$_{17}$ orders at approximately $k_BJ/3\sim$850~K$/3\sim$283~K \cite{lucinski98}. 
Assuming a relative increase of the anisotropy of 10\% ($\Delta K/K =$0.1) and $J/K\sim 10^5$ as in the hybrids, it is predicted from equation \ref{eq:tc2} an increase of the critical temperature of 8.5~K. 
This device would work perfectly in the room temperature regime, where the critical temperature can be optimally tuned by the layer thickness. 
Additionally the spacer can be selected as a function of different external stimuli, 
the only condition being a sufficient modification of the anisotropy $K$. 

In conclusion the magnetization dynamics in both MACuCl and EACuCl can be fully described by a photo-induced enhancement of the axial anisotropy, which causes an increase of the magnetization precession amplitude in MACuCl close to $T_C$. 
The mechanism described in this paper is applicable also to 
multilayer structures consisting of a combination of piezoelectric layers and magnetic layers with a strain dependent anisotropy, in this way we expect that on can even design devices functional at room temperature.

We acknowledge Prof.Dr. A.V.~Kimel, D.~Bossini and Dr.~R.I.~Tobey for useful discussion.\\

%

\end{document}